\documentstyle[12pt,psfig]{article}
\begin{document}
\title{Non-Equilibrium Quantum Electrodynamics}\author{C. Anastopoulos ${}^*$  and A. Zoupas ${}^{\dagger}$  \\ Theoretical Physics Group \\
The Blackett Lab. \\ SW7 2BZ London \\E-mail: can@ic.ac.uk and zoupas@ic.ac.uk
\\ PACS numbers: 03.70.+k, 05.70.Ln, 03.65.Bz \\ Imperial TP/96-97/68}
\date{September 1997}
\maketitle
\begin{abstract}
We employ the influence functional technique to trace out the photonic contribution from full quantum electrodynamics. The reduced density matrix propagator then is constructed for the electron field. We  discuss the role of time-dependent renormalization in the propagator and focus our attention to the possibility of obtaining a dynamically induced superselection rule. As a final application of our formalism we derive the master equation for the case of the field being in an one-particle state in the non-relativistic regime and discuss whether  electromagnetic vacuum fluctuations are not sufficient to produce decoherence in a position basis. 
\end{abstract}
 {\it $*$ Current address: Departament de Fisica Fonamental, Universitat de Barcelona, Av. Diagonal 647, 08028 Barcelona. \\e-mail:charis@ffn.ub.es } \\
{\it $\dagger$Current address: Dep. of Physics, University of Chicago, 5640 S. Ellis Avenue, Chicago, Illinois 60637-1433. \\e-mail: azoupas@rainbow.uchicago.edu}
 \renewcommand {\theequation}{\thesection.\arabic{equation}} 
\let \ssection = \section
\renewcommand{\section}{\setcounter{equation}{0} \ssection}
\pagebreak
\section{Introduction}
\par
Quantum field theory provides a broad  framework 
within which problems in diverse branches of physics
can be formulated and addressed. Its success is primarily based on its inherent flexibility: 
its roots lie within both Hamiltonian and statistical mechanics and this double aspect can be used 
invariably when dealing with particular physical phenomena.
\par
Its large domain of applicability in conjunction with the increasing interest in 
non-equilibrium phenomena has led in recent years into the adoption of successful techniques from 
standard non-equilibrium statistical mechanics or the
quantum theory of open systems 
and their upgrading to fit the field theoretic contexts. 
Hence the Feynman-Vernon influence functional technique, 
 the Schwinger- Keldysh closed time-path formalism ( for extensive bibliography see reference \cite{Hu})
and Zwanzig's projection method \cite{An1,An2} 
have been used to deal with diverse issues including early universe cosmology, 
black hole statistical mechanics condensed matter physics and quantum optics.
\par
The earliest specimen among field theories is of course quantum electrodynamics (QED). 
Being a theory of photons and electrons, it can be used to describe most of the phenomena in ``ordinary'' 
low energy matter. Hence, within its range of applications many issues connected with non-equilibrium phenomena arise. 
In this context one of the older 
(dating back to Lorentz for the case of classical electromagnetism)
 interesting questions is what the effective dynamics of charged particles when its interaction with its own electromagnetic field is taken into account. Or stating it differently, 
how  the vacuum fluctuations of the electromagnetic field affect  the evolution of an electron. 
\par
More than that, the program of decoherence raises important questions
relevant to the domain of QED. It has been proposed that the
superselection rules for conserved charges could be another instant of the 
environment induced superselection rules \cite {Zur}: that is, the 
electromagnetic field considered as an environment makes  superpositions of 
states with different charge rapidly lose their coherence. 
Arguments in favor and against this proposal can be found in  \cite{GKZ, Har}. 
\par
 Also of equal importance is whether single electron states exhibit decoherence in position by virtue of their interaction with the photon vacuum \cite{BaCa, Ford}. The two-slit experiment suggests that this is not the case, but we would like to see in more detail the truth of this assertion.
\par
These have been our physical motivations, but were not the only ones. 
We feel that the use of the influence functional technique 
in field 
theoretic situations has been rather less frequent than the strength of the 
method would require. One reason for that 
might be the fact that it gives a description of the system in a 
Schr\"oddinger picture, which 
has been rather rare in most field theoretic applications. Issues like 
renormalization (which 
is explicitly time-dependent) seem to be  more complicated in this picture, 
Lorentz covariance
is not manifest and calculations are definitely messier.
\par
Compensating for these problems are two important benefits: \\
1) Using the influence functional technique, we can more easily see the effects of the initial condition 
of the environment in the evolution equations. In particular, the effects of the unphysical factorizing 
initial condition ( which is also used  in the other techniques) are more easily understood
and as we argue in section 3 we can find regimes where they can be removed by a renormalization process.
\\
2) The main object in the influence functional formalism is the reduced density matrix propagator, the 
knowledge of which enables us to construct master equations, study the classical limit, decoherence 
phenomena and pass from a field description to a particle one.
\par
From the above discussion, we believe that our aim in this paper has been made clear. We intend to use the influence
functional technique to trace out the effects of the electromagnetic field in full QED, hence obtaining 
the reduced density matrix propagator. This is meant to be our tool for examining the possibility 
of a dynamical origin of the charge superselection rules. As a special application we shall construct from this 
the evolution equation for the regime where the initial field state lies in the one - particle sector, aiming 
to obtain a master equation for a single electron as it interacts with the vacuum fluctuations of 
the photon field.
\par
In the next section we give a brief review on the influence functional formalism,  fixing primarily  the 
notation and introducing some  important objects. The calculation of the propagator for the free 
spinor field with external sources (of great importance for perturbation theory) 
is also briefly sketched. Section 3 is the largest and from the technical side the most important one.
Herein the reduced density matrix propagator for the electron field is calculated, using perturbation theory.
The section includes a discussion of the factorizing initial condition, how one can remove the 
inevitable divergences through a renormalization procedure and the implications of our result for dynamically 
induced superselection rules. In section 4 the special case of a single non-relativistic particle 
is studied. A master equation is obtained and the question of whether the photon vacuum can play the role of a decohering environment is studied in detail.
\par
Before proceeding any further we should mention that a master equation in the 
context of quantum electrdynamics for the reduced dynamics of the charges
has been derived before \cite{Diosi}. It is, to the best of our knowledge, the
first attempt to derive such an equation in QED but it does not utilize the 
functional Schr\"odinger picture and it focuses on different issues.
\section{The model}
Our starting point is the QED Lagrangian
\begin{eqnarray}
\lefteqn{{\cal L} = {\cal L}_e + {\cal L}_{ph} +{\cal L}_{int} = }
\nonumber \\
&&\bar{\psi} (i \gamma^{\mu} \partial_m - m) \psi - \frac{1}{4} F_{\mu \nu} F^{\mu \nu}
+ e \bar{\psi} \gamma^{\mu} A_{\mu} \psi
\end{eqnarray}
We are going to use the influence functional technique to trace out the photon field and derive the master 
equation for the
 reduced dynamics of the electron field. To do this we shall work in the Bargmann representation (considering essentially 
coherent state path integrals) 
 and under the Schr\"oddinger picture.
For more details in the various elements we employ in our formalism the reader is referred to \cite{Kla, Kib, Ber}. The relevant object is the evolution 
operator $e^{-iHt}$, the matrix elements of which can be read from its path-integral expression
\begin{equation}
\langle \bar{\psi}_f,A_f| e^{-iHt}| \psi_i, A_i \rangle = \int [d \bar{\psi}] [d \psi] 
[dA] e^{i S[\bar{\psi}, \psi, A]}
\end{equation}
where the sum is over fields such that $\psi(0) = \psi_i$, $\bar{\psi}(t) = \bar{\psi}_f$, $A^{\mu}(0) = A^{\mu}_i$ 
and $A^{\mu}(t) = A^{\mu}_f$. The knowledge of this object is sufficient to determine the reduced density matrix at 
time $t$
\begin{equation}
\rho(t) = Tr_{ph} \left( e^{-iHt} \rho_0 e^{iHt} \right)
\end{equation}
where the trace is over the photon field degrees of freedom. The reduced density matrix propagator $J$ is defined by
\begin{equation} 
\rho_t(\bar{\psi}_f, \psi'_ f) = \int D\bar{\psi}'_i D \psi_i J(\bar{\psi}_f, \psi'_f;t|\bar{\psi}'_i, \psi_i;0) \rho_0 (\bar{\psi}'_i,\psi_i) 
\end{equation}
and can be written in a path integral form
\begin{eqnarray}
\lefteqn{J(\bar{\psi}_f, \psi_f;t|\bar{\psi}_i, \psi_i;0) =} 
\nonumber
\\
&&\int [d \bar{\psi}] [d \psi] \int [d \bar{\psi}'][ d \psi'] 
\exp\bigg\{i S_{e}[\bar{\psi},\psi] - i S_e[\bar{\psi}',\psi']\bigg \}
W[\bar{\psi},\psi;\bar{\psi}',\psi'] \hspace{0.2in}
\end{eqnarray}
with integration over paths $\bar{\psi}(t) = \bar{\psi}_f$, $\psi(0) = \psi_i$, $\bar{\psi}'(0) = \bar{\psi}_i$ and $\psi'(t) 
= \psi'_f$. 
\footnote{ We denote the path-integration measure (over spacetime fields) by $[d .]$ and the measure over fields 
at one Cauchy surface (single moment of time ) as $D $; the former is a formal expression while the latter is 
the Gaussian measure with respect to which the field Hilbert space is defined.} 
 Here, $W$ is the influence functional containing the information of the dynamics and initial state of the 
photon field. In the case of factorisable initial conditions $\rho_0 = \rho_{e0} \otimes \rho_{ph0}$ this has a path integral expression
\begin{eqnarray}
W[\bar{\psi},\psi;\bar{\psi}',\psi'] = \int DA_f DA_i DA'_i
\int [dA] [dA']\exp\bigg \{iS_{ph}[A] - i S_{ph}[A']\bigg\} \times
\nonumber 
\\
 \exp\bigg \{ i S_{int}[A,\bar{\psi},\psi] - i S_{int}
[A',\bar{\psi}',\psi']\bigg \} \times \rho_{ph0}(A_i,A'_i)\hspace{0.2in}
\end{eqnarray}
with path integration over vector fields such that $A(0) = A_i$, $A'(0) = A'_i$ and $A(t) = A'(t) = A_f$.
Since the Lagrangian contains an interaction term of third order to the fields,
 we have to rely on perturbation theory. For this it is sufficient that we compute the 
spinor propagator with external sources.
\subsection{The spinor propagator with external sources}
Let us consider the Lagrangian density
\begin{equation}
{\cal L} = \bar{\psi} (i \gamma^{\mu} \partial^{\mu} - m) \psi + \bar{\eta} \psi + \bar{\psi} \eta
\end{equation}
in terms of the Grassmann fields $\psi$, $\bar{\psi}$ and sources $\eta$, $\bar{\eta}$. Written in terms of the Fourier transformed fields and sources
\begin{eqnarray}
\psi = \sum_p y_p e^{-ip x} \quad \quad \psi^{\dagger} = \sum_p y_p^{\dagger} e^{i p  x} \\
\eta = \sum_p  \eta_p e^{-ip x} \quad \quad \psi^{\dagger} = \sum_p \eta_p^{\dagger} e^{i p  x} \end{eqnarray}
it reads
\begin{equation}
L = \int d^3x {\cal L} = \sum_p y_p^{\dagger} ( i \partial_0 - i \gamma^0 \gamma^i p_i -m \gamma^0)y_p +
\eta_p^{\dagger} \gamma^0 y_p + y^{\dagger}_p \gamma^0 \eta_p
\end{equation}
which corresponds to a Hamiltonian 
\begin{equation}
H = \sum_p y_p^{\dagger} ( \gamma^0 \gamma^i p_i + m \gamma^0) y_p - \eta^{\dagger}_p \gamma^0 y_p - y^{\dagger} \gamma^0 \eta_p
\end{equation}
Now, $y_p$ is a four-component Grassmann variable. In order to write the Hamiltonian as a sum of 
single-component Grassmann variables one has to diagonalise the matrix
\begin{equation}
\gamma^0 \gamma^i p_i + m \gamma^0
\end{equation}
Writing in the standard way (suppress the index $p$) $y = y_i u_i + z_i v_i$, $\eta = \eta_i u_i + \eta'_i v_i$,
$ i =1,2$, there is a particular choice of $u$,$v$ such that the above matrix becomes diagonal. 
The Hamiltonian becomes therefore a sum of terms of the form
\begin{equation}
H = \omega \bar{y} y - \bar{\eta}y - \bar{y} \eta
\end{equation}
where $\omega_ p^2 = ({\bf p}^2 + m^2)$.
This is just the Hamiltonian of a Grassmann harmonic oscillator with external sources, the propagator of which reads in a path-integral form
\begin{equation}
\langle \bar{y}_f, t | y_i,0 \rangle = \int Dy D \bar{y} e ^{i S[y,\bar{y}]}
\end{equation}
where summation is over paths with $y(0) = y_i$, $\bar{y}(t) = \bar{y}_f$ and the action is 
\begin{equation}
i S[y,\bar{y}] = \bar{y}y(t) +i \int_0^t ds (i \bar{y} \dot{y} - H(y, \bar{y}))
\end{equation}
The path integral  integral is readily evaluated by the saddle-point method to yield
\begin{equation}
\langle \bar{y}_f,t|y_i,0 \rangle = e^{i S_{cl}}
\end{equation}
with
\begin{eqnarray}
\lefteqn{i S_{cl} =\bar{y}_f y_i e^{-i \omega t} + i \int_0^t ds \big[ e^{-i\omega(t-s)} \bar{y}_f \eta(s) +
e^{-i \omega s} \bar{\eta}(s) y_i \big]} \hspace{0.9in}
 \nonumber 
\\
&&-\int_0^t ds \int_0^s ds' e^{-i \omega |s-s'|}\; \bar{\eta}(s) \eta(s)
\end{eqnarray}
One can thus use the inverse Fourier transform to construct the propagator for the full spinor field 
\begin{eqnarray}
\lefteqn{\langle \bar{\psi}_f,t| \psi_i,0 \rangle_{\eta \bar{\eta}} = 
\exp \Bigg\{ \int d^3x d^3x'\Bigg[ \bar{\psi}_f(x) \Delta(x,x';t) 
\psi_i(x')} \hspace{0.3in} 
\nonumber 
\\
&&+ i \int_0^t ds \bigg(\bar{\psi}_f(x) \Delta(x,x';t-s)
\eta(x',s) + \bar{\eta}(x,s) \Delta(x,x';s) \psi_i(x')\bigg) \hspace{0.3in}
\nonumber 
\\
&&-\int_0^t ds \int_0^{s'} \bar{\eta}(x,s) \Delta(x,x';|s-s'|)\eta(x',t) 
\Bigg] \Bigg\}\hspace{0.3in}
\end{eqnarray}
where $\Delta$ is a ``propagating kernel'' 
\begin{equation}
\Delta(x,x';t) = \int \frac{d^3p}{(2 \pi)^3} e^{-i p  (x-x')} 
\left( \frac{\omega_p \gamma^0 
- \gamma^i p_i}{m} \cos \omega_p t - i \sin \omega_p t\right)
\end{equation}
\footnote{$\Delta$ is essentially a unitary  operator in the single-particle Hilbert space out of which the  field 
Fock  space is constructed.} 
\section{The reduced density matrix propagator}
\par
Let us concentrate now on the photon  field dynamics. The relevant part of the Lagrangian density is
\begin{equation}
{\cal L} = -\frac{1}{4} F_{\mu \nu} F^{\mu \nu} + J_{\mu} A^{\mu}
\end{equation}
where $J^{\mu} = \bar{\psi} \gamma^{\mu} \psi$
\par
The main idea in the influence functional formalism is to proceed and use the propagator for the above Lagrangian ,
while treating $J$ as an external; source. Now this is clearly a linear action, hence the propagator in presence of sources can be computed explicitly.
\par
It is more convenient to fix the gauge for the photon field ($\partial_{\mu} A^{\mu} = 0$, $A^0 = 0$ and expand the 
spatial components of the field in normal modes
\begin{equation}
A^i(x) = \sum_k \sum_r \epsilon^i_{(r)}({\bf k}) \left( q_{1r}({\bf k}) \cos{\bf kx} + q_{2r}^i({\bf k}) \sin {\bf kx} \right)
\end{equation}
where the polarization matrices satisfy
\begin{equation}
\sum_r \epsilon^i{(r)} ({\bf k}) \epsilon^j{(r)} ({\bf k}) = T^{ij}({\bf k}) = \delta^{ij} - k^i k^j /{\bf k}^2
\end{equation}
Substituting this into (2.6) we get the following expression for the relevant part of the action
\begin{equation}
S_{ph} + S_{int} = \frac{1}{2} \int_0^t ds \sum_k \sum_r \sum_l  (\dot{q}^2_{rl}({\bf k}) + k^2 q^2_{rl}({\bf k})
 + 2 q_{rl}({\bf k}) J_{rl}({\bf k})
\end{equation}
In this expression the index $l$ takes values $1$ and $2$ according to the decomposition (3.2) and the external fields $J$ read
\begin{eqnarray}
J_{1r} ({\bf k}) = e \int d^3x \cos{\bf kx} J^i(x) \epsilon_{i(r)}({\bf k}) \\
J_{2r} ({\bf k}) = e \int d^3x \sin{\bf kx} J^i(x) \epsilon_{i(r)}({\bf k}) 
\end{eqnarray}
The action corresponds to an infinite number of harmonic oscillators under external source and the propagator can be computed the standard way. Moreover if we assume the initial state factorisable and the photon part to correspond to the vacuum $\rho_{ph} =|0 \rangle \langle 0 |$ we can use equation (2.19) to compute the influence functional. It is a standard calculation to show that
\begin{eqnarray}
W[J,J'] = C(t) \prod_{\alpha} \exp \left[ -i \int_0^t ds_1 \int_0^{s_1} ds_2 [J_{\alpha} - J'_{\alpha}] (s_1)
\right. \nonumber \\
\left. \times \left( \frac{e^2}{2 \omega_k} \sin \omega_k (s_1 - s_2) \right) [J_{\alpha} + J_{\alpha}'](s_2) \right.
 \nonumber \\
\left. -  \int_0^t ds_1 \int_0^{s_1} ds_2 [J_{\alpha} - J'_{\alpha}] (s_1) 
\left( \frac{e^2}{2 \omega_k} \cos \omega_k (s_1 - s_2) \right) [J_{\alpha} - J_{\alpha}'](s_2) \right]
\end{eqnarray}
where $\alpha$ is a collective index containing ${\bf k}$, $l$ and $r$.
Using the inverse Fourier transform we obtain the final result for the influence functional
\begin{eqnarray}
W[J,J'] = C(t) \exp \left[\delta_{ij }\int d^3x d^3x'\int_0^t ds_1  \int_0^{s_1} ds_2  \right. \nonumber \\
\left. \left(  -i[J^i - J'^i](x,s_1) \eta_{ij}(s_1 - s_2,x,x´)
[J^j + J'^j](x',s_2) \right. \right. \nonumber \\
\left. \left.
 -[J^i - J'_i](x,s_1) \nu_{ij}(s_1 - s_2,x,x´)
[J_j - J'j](x',s_2) \right) \right]
\end{eqnarray}
Here, $\eta$ and $\nu$ are the dissipation and noise kernel respectively which read
\begin{eqnarray}
\eta_{ij}(s,x,x´) = e^2 \int \frac{d^3k}{2 \omega_k (2 \pi)^3} e^{-i{\bf k}(x-x´)} T^{ij}({\bf k}) \sin \omega_k s \\
\nu_{ij}(s,x,x´) = e^2 \int \frac{d^3k}{2 \omega_k (2 \pi)^3} e^{-i{\bf k}(x-x´)} T^{ij}({\bf k}) \cos \omega_k s
\end{eqnarray}
\par
Having the expression for the influence functional we can proceed to evaluate 
the reduced density matrix propagator. But first we should recall that its 
derivation depended upon the assumption of factorised  initial state. We 
should first then discuss what is the physical meaning of such a condition.

\subsection{ The initial condition}
Our choice of the initial condition (factorisable density matrix, vacuum for 
the photon field) has been such as to make the calculations easier, but on 
the other hand it introduces a number of problems . This condition has been 
extensively used in quantum Brownian motion models and is generally deemed 
unphysical. Such a separation between system and subsystem requires a large 
amount of energy (in the case  of fields an infinite amount) and hence 
cannot be taken as realistic.  The effects of this initial condition have
not been fully understood yet.  It was conjectured \cite{HPZ} that it is
related to a jolt in the diffusion coefficients at early times, dissapearing
in a timescale proportional to $\Lambda^{-1}$, where $\Lambda$ is the high 
frequency cut-off of the environment.  Nevertheless this is not probably the
case. Recent studies \cite{RP} have shown that this jolt is present even if 
one 
assumes correlated initial states. The typical behavior in those systems
is that the preferred degrees of freedom at early times couple very fast to 
the high frequencies of the environment, until a dynamical equilibrium is 
established and then at later times the low frequency modes start playing an 
important effect (in essence the long time limit is governed solely by the 
latter). In this class of systems it is usually  the case that
$\Lambda \gg \omega$ where $\omega$ is the typical frequency of the preferred 
degrees of freedom.  To conclude, it is the existence of the 
above dynamical equilibrium allowing us to trust the analysis with the 
given initial condition. 
\par
Doing field theory just makes things more difficult as one would expect. 
In QBM models the separation between system and environment follows naturally
by assuming the typical timescales of the problem and somehow the environment 
appears to be ``robust'' to the relevant degrees of freedom. Backreaction 
effects need not be taken into account as far as the environment is concerned.
One of the concequencies is that a cut-off naturally appears only in the 
frequencies related to the environmental degrees of freedom ($\Lambda$). 
In QED this separation is not obvious. We could equally 
well formulate the inverse problem {\it i.e.} what happens to the photon
field if we trace out the fermionic degrees of freedom?  This, for
example, is a question of relevance to solid state physics.
Still we expect that some of the features of QBM models will be relevant but 
we have to be very careful in our analysis.
\par
Our aim is to find the regime where our initial condition can be a good 
approximation, in the sense that a suitable renormalization of 
the parameters can be expected to give reliable results.  What we assume is 
that at times $t < 0$ the  electron - photon field combined system
lies on its ground state and at time $ t = 0$ an operation is performed on 
the spinor sector, which effects to a local excitation 
of spinor degrees of freedom. Our aim is to see how the reduced density 
matrix describing these degrees of freedom evolves in time. The vacuum of 
the full QED theory is a correlated one, so no operation carrying finite 
energy is sufficient to fully separate the spinor from the electromagnetic 
degrees of freedom. There will be correlations stemming from the ultra-
violet and infra-red sector of the photon spectrum.
\par 
Now, in perturbative quantum field theory, one is starting from the Hilbert
space of free fields (hence considering a factorised vacuum state) and
then the renormalization procedure provides a mapping (though a mathemati
cally ill-defined one
\footnote{ Since the free and interacting field Hilbert space carry 
unitarily inequivalent representations of the
canonical (anti)commutation relations}) from the free fields Hilbert space 
to the interacting one \cite{Ber, Haag}.
This observation suggests a  rephrasing of our problem  : is there any 
renormalization 
procedure enabling us to remove the effects of the initial condition? 
\par
The expected behavior is the following. At early times the high frequency 
modes of the photon field couple very fast to the slow ones of the fermion
field and vice-versa.  This latter coupling expresses the backreaction of the 
relevant degrees of freedom to the environment (We believe backreaction
effects to be important as far as decoherence is concerned). Our experience 
tells as that it will only affect the initial moments of the dynamics of the 
photon field had we wished to study them. Hence in the times where the 
problems of the unphysical initial condition are expected to arise the 
contribution of correlations with soft photon modes is not important. This 
leads us to expect that the standard renormalization procedure of the 
propagator \cite{Ber} must be sufficient to account for the effect of 
the soft modes. This will be verified in our analysis, that no divergent 
contributions stemming from low energy modes appear in the density matrix 
propagator apart from the standard ones due to perturbative evaluation.
\par
The case is more complicated for the highly energetic modes, since there 
exists an infinite energy barrier between our factorised initial state and a 
physical one ({\it i.e.} the Hilbert space of the interacting theory does not 
contain a factorized state). 
This implies that the time it will take for these modes to come to equilibrium
will tend to become infinite. To get over this difficulty we may observe that
no such modes can be excited with energy larger than the total energy of the 
system. Hence there exists of a natural cut-off $\Lambda$ (state-dependent) 
for both spinor and photon modes (This is due to the fact that photon and 
fermion fields involve typical frequencies which 
are of comparable order of magnitude. There is not a 
preferred time scale distinguishing one from the other in a natural way).
Introducing this cut-off can be taken as considering factorised initial state 
for only the modes with energy less 
than $\Lambda$, while higher energetic ones (for both fields) can be taken to 
be in dynamical equilibrium \footnote{ More precisely we can assume a 
truncation of the field theory such that high energy modes do not contribute 
in the dynamics and hence lie close to their vacuum state}. This implies that 
the effect of the spurious initial condition is to be contained in the parts 
of the propagator containing the cut-off. Hence a renormalization procedure 
(additional to the standard dynamical one) removing these terms would yield 
reliable results for sufficiently large times.
\par
More formally, one could write that the physical density matrix should be
\begin{equation}
\rho_0 =  \rho_e \otimes \rho_{ph} + e^2 \rho_{cor}
\end{equation}
The first term contains a factorized state for all modes it describes while 
the second one gives the contribution of the correlated vacuum of the higher
modes. In other words we are asking for a state which in the low energy
modes is essentially uncorrelated while in the high energy ones contains the
correlations of the real vacuum state of the interacting theory.
The later contains differences from the factorized vacuum of the order of
$e^{2}$. Time evolution to lowest order in perturbation theory takes the low
energy modes to low ones and similarly for the high ones. To have mixing 
of the regimes we must consider orders $e^{2}$ . This means that the 
contribution of the initially correlated terms (already of order $e^{2}$)
to the low energy ones at late times will be of order $e^{4}$, hence we 
can safely ignore it in the lowest order of perturbation expansion.
\par
To summarize: there is a regime of energies $E$, where a (time-dependent) 
renormalization on the propagator 
is expected to remove all effects due to our unphysical initial condition and 
our expressions can be valid for times $t >> E^{-1}$. 
In particular, the non-relativistic limit for a single particle where 
$ E \simeq m$ should  definitely lie within our range of validity.

\subsection{ The propagator}
\paragraph{The index notation}
It is more convenient to use an index notation for the spinor fields so that we will not have to carry the 
explicit dependence on $x$.
We have essentially four types of spinor fields: $\psi(x)$, $\bar{\psi}(x)$ , $\psi'(x)$ ,$\bar{\psi}'(x)$. 
The first two propagate forward in time and the other backwards (note the similarity with the CTP formalism). Now, $\psi$ and $\bar{\psi}'$ in the path integration are fixed by their initial condition at $t = 0$; we shall assume  
that they carry an upper index (in place of their x dependence), while $\bar{\psi}$ and
$\psi'$ are fixed by their values at $t$ and we shall use a lower index for them. This is an elaboration of the notation used in \cite{An2}.
\par
Contraction of indices amounts to integration with respect to $x$, while kernels carry indices 
according to the fields they are contracted. Hermitian conjugation for a kernel amounts to inversion of all indices. 
Hence for example our propagator (2.19) reads in this notation
\begin{eqnarray}
\langle \bar{\psi}_f,t| \psi_i,0) = \exp \left[ \bar{\psi}_{fa} \Delta(t)^a_b \psi_i^b + i \int_0^t ds (\bar{\psi}_{fa} \Delta^a_b(t-s)
\eta^b(s) + \right. \nonumber \\
\left. \bar{\eta}_a(s) \Delta^a_b(s) \psi_i^b) - \int_0^t ds \int _0^s ds' \bar{\eta}_a(s) \Delta^a_b(|s-s'|) \eta^b(s') \right]
\end{eqnarray} 
Whenever an upper and lower index do not denote integration and are to be thought as free a bar will appear in top of them.
\par
\paragraph{Perturbation expansion}
To compute the propagator one needs to evaluate the path integral (3.8) using the expression (2.6) 
for the influence  functional. Since the influence functional contains term of fourth order to the fields, 
an exact calculation is infeasible, hence we will have to rely on perturbation theory. 
For this we shall employ the following identity
\begin{eqnarray}
J(\bar{\psi}_f,\psi'_f;t|\bar{\psi}'_i, \psi_i;0) = 
W[ -i \frac{\delta}{\delta \eta}, i \frac{\delta}{\delta \bar{\eta}}; i \frac{\delta}{\delta \eta'}, -i \frac{\delta}{\delta \bar{\eta}'} ]   \nonumber \\
\langle \bar{\psi}_f,t| \psi_i,0 \rangle_{\eta \bar{\eta}} \langle \psi'_f,t| \bar{\psi}'_i,0 \rangle_{\eta' \bar{\eta'}} |_{\eta = \bar{\eta} = \eta' = \bar{\eta}' = 0}
\end{eqnarray}
Expanding $W$ in powers of $e^2$ one can use this formula to derive the perturbative series for 
the density matrix propagator. 
\par
Now, in the expansion we shall encounter loop terms (corresponding to $Tr \Delta(0)$ or 
$\Delta^{a}{}_{b}(t-s) \Delta^{b}{}_{a}(s)$. 
Their effect would be to  produce a change on the kernel $\Delta$ and to necessitate a field and coupling 
constant renormalization . We will prefer to work in the zeroth loop order, that is consider only tree diagrams
\footnote{We could obtain the same result had we performed a saddle point approximation in the path integral and used perturbation theory in order to construct the corresponding ``classical'' equations of motion.} .
\par
Hence, when taking only the tree diagrams into consideration we see that in the propagator there appear exponentiated four interacting terms.
\begin{eqnarray}
&&-\int_0^t ds \int_0^s ds' \bigg[ \bar{\psi}_{fa}\Delta^a{}_{\bar{e}}(t-s)\gamma^i 
\Delta^{\bar{e}}{}_{b}(s) \psi^b_i \mu_{ij,eg} (s-s')
\nonumber 
\\ 
&&\hspace{2in}\bar{\psi}_{fc} \Delta^c{}_{\bar {g}}(t-s') \gamma_j 
\Delta^{\bar{g}}{}_d (s') \psi_i^d\bigg]
\nonumber 
\\
&&+\int_0^t ds \int_0^s ds' \bigg[ \bar{\psi}_{fa}\Delta^a{}_{\bar{e}}(t-s)\gamma^i 
\Delta^{\bar{e}}{}_b \psi_{i}^{b}(s)\mu^{*}_{ij,eg}(s-s')
\nonumber
\\
&&\hspace{2in}\bar{\psi'}_{i}^c \Delta_c{}^{\bar{g}}(s') \gamma^j \Delta_{\bar{g}}{}^d (t - s')
\psi_{fd}' \bigg]
\nonumber 
\\
&&+\int_0^t ds \int_0^s ds' \bigg[ \bar{\psi'}_{i}^a\Delta_a{}^{\bar{e}}(s)\gamma^i  
\Delta_{\bar{e}}{}^b(t-s) \psi_{fb}' \mu_{ij,eg}(s-s')
 \nonumber 
\\
&&\hspace{2in}\bar{\psi}_{fc} \Delta^c{}_{\bar{g}}(t-s') \gamma^j
\Delta^{\bar{g}}{}_d (s') \psi_i^d \bigg]
\nonumber
\\
&&-\int_0^t ds \int_0^s ds' \bigg[ \bar{\psi'}_{i}^{a}
\Delta_a{}^{\bar{e}}(s)\gamma^i \Delta_{\bar{e}}{}^b(t-s) \psi'_{fb} \mu^{*}_{ij,eg}(s-s')
 \nonumber 
\\
&&\hspace{1.9in} \bar{\psi'}_{i}^{c} \Delta_{c}{}^{\bar{g}}(s') \gamma^j 
\Delta_{\bar{g}}{}^d (t-s')\psi'_{fd} \bigg]
\end{eqnarray}
In the above expression $\mu(s-s')$ stands for
$(\nu+i\eta)(s-s')$ and $\mu^{*}$ denotes its complex conjugate.  
We observe in these terms combinations of the form $\Delta \gamma \Delta$, which amount to products of three and two $\gamma$ matrices. For physical interpretation it would be more convenient to write these in terms of a fixed basis of the $4 \times 4$ vector spaces where the Grassmann variables. Actually in our case (parity preservation) only $\gamma^{\mu}$, $1$ and
$\sigma^{\mu \nu}$ are relevant. One can use the identities
\begin{eqnarray}
\gamma^{\mu} \gamma^{\nu} &=& \eta^{\mu \nu} 1 + i \sigma^{\mu \nu} 
\\
\gamma^{\mu} \gamma^{\nu} \gamma^{\rho} &=& g^{\mu \nu} \gamma^{\rho}
+ g^{\nu \rho} \gamma^{\mu} - g^{\mu \rho} \gamma_{\nu} 
\end{eqnarray}
to bring all interaction terms into sums of the form $(\bar{\psi} A
\psi)(\bar{\psi} B \psi)$ where $A$ and $B$ are matrices:  $1$,$\gamma^{\mu}$, $\sigma^{\mu \nu}$. 
\par
Let us denote by capital indices $I$,$J$, \ldots whether the indices
correspond to scalar $S$, vector  $V$, tensor $T$ current (in terms of Lorentz indices $S$ stands for no index , 
$V$ for a single index and $T$ for two antisymmetric indices). With this notation we can write the reduced density 
matrix propagator as
\begin{eqnarray}
\lefteqn{J(\bar{\psi}_f, \psi'_f;t| \psi_i, \bar{\psi}_i',0) = }
\nonumber
\\
&&\exp \bigg( \bar{\psi}_{fa} \Delta^a{}_{b} \psi_i^b + 
\bar{\psi}'^a_i \Delta_a{}^b \psi'_{fb}+ 
\epsilon_{_{IJ}}{}^{a}{}_{b}{}^{c}{}_{d} \bar{\psi}_{fa} A_I \psi_i^b
\bar{\psi}_{fc}A_J \psi_i^d +
\nonumber 
\\
&&\epsilon_{_{IJ}}{}_{a}{}^{b}{}_{c}{}^{d} 
\bar{\psi'}_i^a A_I \psi'_{fb} \bar{\psi'}_i^c A_J \psi'_{fd}- 
\zeta_{IJ}{}^{a}{}_{bc}{}^{d} 
\bar{\psi}_{ia}A_I \psi_f^b \bar{\psi'}_i^c A_J \psi'_{fd} \bigg)
\end{eqnarray}
The coefficients $\epsilon$ and $\zeta $ can be explicitly
computed. Their Fourier transforms are
\begin{eqnarray}
\epsilon_{IJ}(x,x',y,y') = \int \frac{d^3p}{(2 \pi)^3} \frac{d^3p'}{(2 \pi)^3} e^{_ip(x-x') - ip'(y-y')}
\epsilon_{IJ} (p,p',t)\\
 \zeta_{IJ}(x,x',y,y') = \int \frac{d^3p}{(2 \pi)^3} \frac{d^3p'}{(2 \pi)^3} e^{_ip(x-x') - ip'(y-y')}
\zeta_{IJ} (p,p',t)
\end{eqnarray}
\paragraph{Renormalization}
Divergences appear into the coefficients of the propagator. We have chosen a naive reqularisation scheme of 
substituting $\int d^3k $ with $ \int d^3k e^{-k \epsilon}$. Divergences then take the form of either poles 
in $\epsilon$ or logarithms of $\epsilon$. The renormalization procedure we are going to follow consists in 
essentially dropping all terms containing the divergences. There is an ambiguity as far as the logarithmic 
divergences go, in the sense that one can with a simple rescaling of $\epsilon$ terms with no dependence on 
the cut-off appear. Happily, in the case of such divergences the ambiguous terms are  oscillatory in time and 
generically negligible at late times, when the approximation is thought to be valid.

\subsection{The master equation}
From the form (3.17) of the reduced density matrix propagator it is straightforward to obtain the master equation.
Recall that the relevant procedure consists  in writing all multiplicational action of initial variables operators
on $J$ to differential action of final variables. It can be easily shown that the master equation is of the form
\begin{eqnarray}
\lefteqn{\frac{\partial}{\partial t} \hat{\rho} =
\frac{1}{i}[\hat{H},\hat{\rho}] }
\nonumber
\\
&&\hspace{0.2in}+ \int \frac{ d^3 {\bf p}}{(2 \pi)^3} \frac{ d^3 {\bf p'}}{(2 \pi)^3}  
\frac{ d^3 {\bf k}}{(2 \pi)^3} 
\bigg(i \; e^2 \alpha_{_{IJ}}({\bf p}, {\bf p'},{\bf k};t) 
\big[ \hat{J}_I({\bf p},{\bf k})
\hat{J}_J({\bf p'},{\bf k}), \hat{\rho} \big]\hspace{0.5in}
\nonumber 
\\
&&\hspace{1.6in}- e^2\beta_{IJ}({\bf p}, {\bf p'}, {\bf k};t) \big[\hat{J}_I({\bf p},{\bf k}), 
\big[ \hat{J}_J({\bf p'},{\bf k}), \hat{\rho}\big] \big] \bigg)\hspace{0.5in}
\end{eqnarray}
The kernels $\alpha$ and $\beta$ can be computed from the knowledge of $\epsilon$ and $\zeta$.
\par
It is important to remark that the interaction of the spinor field with its environment comes through terms quadratic in 
the currents $J_I({\bf p},{\bf k}) =  \bar{\psi} A_I \psi ({\bf p}+{\bf k})$. 
\par
What are the consequences of the expression (3.20.) as far as decoherence phenomena for the 
electron field are concerned? The complicated expressions for our coefficients does not support the probability of giving a general argument or proof. What they hint is that if we are to deal with those issues  we must restrict ourselves in particular regimes. Such a regime (of a single particle state in the non-relativistic limit) we shall study in the next section.
\par
Still, based on the experience obtained by study of simpler open systems 
\cite{Zuretal} 
one can make a number of conjectures with high 
plausibility. Both dissipation and diffusion terms in the master equation are expressed in terms of the three currents (scalar, vector, antisymmetric two-tensor). This means that they form the channels through which the spinor field interacts with its environment. In particular, the diffusion term for each current suggests that the reduced density matrix has the tendency to become approximately diagonal in some basis close to  eigenstates of the current operators
 As it stands, we cannot determine which such a basis can be, let alone whether the suppression of the off-diagonal terms is sufficiently strong. But if such regimes exist, then superposition of states with large difference in the expectation values of the current operators would decohere fastly. We could then say that we have some sort of ``local current '' environment-induced superselection rules.  This is a more general case than a single charge superselection rule and actually seems more physical: Charge superselection rule is a special case of this, but we could also have decoherence between states with the same global charge but different local charges, i.e between two seventeen-electron states with expectation values of currents having support in spatial neighborhoods with large separation. The study of such regimes would be invaluable towards understanding the role of decoherence in quantum field theories and can in principle
 be performed starting from the propagator (3.17).  In connection to the above 
it is also worth noticing that not all of the currents appearing in Eqn (3.20) 
are conserved. Only the vector current has got this property. This current is 
thus related to a conserved quantity, a charge, given as a volume integral 
over the corresponding local density. When this volume is taken to infinity it 
will be an exactly conserved quantity.  Then in this case according to 
\cite{Har} we have exact decoherence which because of the conservation law
is related to the existence of an exact superselection rule.  Nothing like
that is  obvious from the form of the above master equation.  It is not very 
clear to us how the results of \cite{Har} could be reproduced in the given
formalism. We believe it to be an issue worth investigating. Another 
interesting point could be made in relation  to the ideas of Gell-Mann and 
Hartle \cite{GeHa} that in decoherent histories hydrodynamic variables are the
variables that habitually decohere, being the quasiclassical variables of
genuinly closed systems. These variables are related to locally conserved 
currents.
Since her  not all of our currents are conserved there is an open question of  
what decoherence in the basis of those currents (if the phenomenon occurs) means
and whether the basis of the conserved current is in any sence prefered over 
the other ones as one would naively expect. 

\section{The master equation for a single particle}
The reduced density matrix propagator for the spinor field (3.17) is quite general, and can act 
as a starting point for exploring particular problems . In this section, we will derive the master
 equation for a single non-relativistic particle, having as our starting point the field propagator.
\par
The general technique for the derivation of master equations for particle dynamics, starting from
field theoretic evolution equations has been developed in reference \cite{An2}, the notation of
which we shall employ. 
In our treatment we are going to make the following assumptions:
\\ 
i) The initial state of the field lies within the one-particle sector.
\\
ii) We work in frame of reference such that the momentum of the initial state lies in the 
non-relativistic regime: $|{\bf p}/m| << 1$. This means in particular that there is no 
pair creation and the master equation is  expected to be trace-preserving.
\\
iii) We work to the lowest non-trivial order in $e^2$ and    $|{\bf p}/m| $.
\par
The assumption (i) guarantees that the propagator of the single particle is of the form
${\bf PJP}$, where ${\bf J}$ is the field propagator and ${\bf P}$  projector on
the space of field states that project to the one-particle sector. It reads explicitly
\begin{equation}
{\bf P} \rho = P_1 \rho P_1
\end{equation}
where the operator $P_1$ is defined by its symbol in the Bargmann representation
\begin{equation}
\tilde{P}_1(\bar{\psi}, \psi) = \bar{\psi} P \psi
\end{equation}
Here $P$ is a $4 \times 4$ matrix that projects into the two dimensional subspace of Dirac
 spinors that correspond to particles. In the Dirac-Fermi  representation for the $\gamma$
matrices it reads $P = diag(1,1,0,0)$.
\par
The propagator $J_{1par}$for the single particle then reads
\begin{eqnarray}
\lefteqn{J_{1par}( \bar{\chi}_f, \chi'_f;t| \bar{\chi}'_i, \chi_i;0)=} 
\hspace{0.3in}
\nonumber
\\
&& \int D\bar{\psi}_f D \psi_f D \bar{\psi}_i D \psi_i 
D \bar{\psi}'_f D \psi'_fD \bar{\psi}'_i D \psi'_i 
\; e^{[- \bar{\psi}_f  \psi_f - \bar{\psi}_i  \psi_i  -\bar{\psi}'_f
\psi'_f -\bar{\psi}'_i  \psi'_i ]}\hspace{0.4in}
\nonumber 
\\  
&&\hspace{0.5in}\times (\bar{\chi}_f  P \psi_f) ( \bar{\psi}'_f P \chi'_f) 
J(\bar{\psi}_f. \psi'_f;t| \bar{\psi}'_i, \psi_i; 0 ) 
(\bar{\chi}'_i P \psi'_i) (\bar{\psi}_i P\chi_i)
\end{eqnarray}
 It is easily shown that $J_{1par}$ is of the form
\begin{equation}
J_{1par}( \bar{\chi}_f, \chi'_f;t| \bar{\chi'}_i, \chi_i;0) =
{K^a{}_b{}_c{}^d} \bar{\chi}_{fa} \
{\chi}_f^{'b} \  \bar{\chi}_i^{'c} \  \chi_{id}
\end{equation}
The kernel ${K^a{}_b{}_c{}^d} = K_{r_{f} r_{i} s_{f} s_{i}}(x_f,y_f;t|x_i,y_i)$ is the single
 particle propagator ( the indices $r$ and $s$ take values $1$ and $2$ and correspond to 
the two spin polarizations). To lowest order in $e^2$ the kernel reads
\begin{eqnarray}
K^a{}_b{}_c{}^d &=& \Delta^a{}_b \Delta_c{}^d 
\nonumber 
\\
&& +\int_0^t ds \int_0^s ds' \bigg[ P \Delta^a{}_{\bar{e}}(t-s)\gamma^i
\Delta^{\bar{e}}{}_b(s)P \mu^{*}_{ij,eg}(s-s')
\nonumber
\\
&& \hspace{1.7in} P\Delta_c{}^{\bar{m}}(s') 
\gamma^j \Delta_{\bar{m}}{}^d (t -s')P \bigg] 
\nonumber 
\\
&& +\int_0^t ds \int_0^s ds' \bigg[ P \Delta^a{}_{\bar{e}}(s)\gamma^i
\Delta^{\bar{e}}{}_b(t-s)P \mu_{ij,eg}(s-s')] 
\nonumber
\\
&& \hspace{1.8in} P\Delta_c{}^{\bar{g}}(t-s') 
\gamma_i\Delta_{\bar{g}}{}^d (s')P  \bigg] 
\end{eqnarray}
The non-relativistic limit is more clearly obtained
in the Dirac-Fermi representation for the $\gamma$ matrices
\begin{eqnarray}
\gamma^0 = \left( \begin{array}{cc} 
                  1&0\\
                  0&-1\\ \end{array} \right)\;\;
\gamma^i = \left( \begin{array}{cc} 
                  0&\sigma^i \\
                  \sigma^i &0 \end{array} \right)
\end{eqnarray}
In this representation the kernel $\Delta$ reads (its Fourier transform)
\begin{eqnarray}
\Delta({\bf p};t) = \left( \begin{array}{cc} 
\frac{\omega_p}{m} \cos\omega_pt - i \sin \omega_pt&
-\frac{\bf{\sigma p}}{m} \sin \omega_pt \\
 -\frac{\bf{\sigma p}}{m} \sin \omega_pt & -\frac{\omega_p}{m} \cos\omega_pt - i \sin \omega_pt
\end{array} \right)
\end{eqnarray}
Substituting this into (4.5) we derive the following expression for the propagator
\begin{eqnarray}
\lefteqn{K_{r_f r_i s_f s_i}(x_f,y_f;t|x_i,y_i) =\int \frac{d^3{\bf p}}{(2 \pi)^3}
\frac{d^3{\bf p'}}{(2 \pi)^3} e^{i \big(p' (x_f - x_i) - p (y_f - y_i) 
-\frac{t}{2m} (p^2 - p'^2 )\big)}}\hspace{1.4in}
\nonumber
\\
&&\times\Bigg[ 1 + \frac{e^2}{m^2} \int \frac{d^3k}{(2 \pi)^3} e^{ik (x_i - y_i)} 
T^{ij}({\bf k}) \int_0^t ds \int_0^s ds'\hspace{0.25in}
\nonumber
\\
&&\times\Lambda^{k}(t,s) ( \sigma^k \sigma^i)_{r_f r_i} e^{-i k (s-s')} (\sigma^j \sigma^l)_{s_f s_i} (\Lambda'^{^{k}}(t,s'))^{*}\hspace{0.25in}
\nonumber 
\\
&&\times (\Lambda^{k}(t,s))^{*} ( \sigma^k \sigma^i)_{s_f s_i} e^{i k (s-s')} (\sigma^j \sigma^l)_{s_f s_i} \Lambda'^{^{l}}(t,s')\Bigg]\hspace{0.25in}
\end{eqnarray}
%
% Old version
%
%\begin{eqnarray}
%\lefteqn{K_{r_f r_i s_f s_i}(x_f,y_f;t|x_i,y_i) =} 
%\nonumber
%\\
%&&\int \frac{d^3{\bf p}}{(2 \pi)^3}
%\frac{d^3{\bf p'}}{(2 \pi)^3} e^{i \big(p' (x_f - x_i) - p (y_f - y_i) 
%-\frac{t}{2m} (p^2 - p'^2 )\big)}
%\;\Bigg[ 1 + \frac{e^2}{m^2} \int \frac{d^3k}{(2 \pi)^3} e^{ik (x_i - y_i)} 
%\nonumber 
%\\ 
%&&\times T^{ij}({\bf k}) \int_0^t ds \int_0^s ds´ 
%( i p^k f(t,s) + k^k g(t,s)) ( \sigma^k \sigma^i)_{r_f r_i} e^{-i k (s-s´)} (\sigma^j \sigma^l)_{s_f s_i}
%\nonumber
%\\
%&&\hspace{3in}(-i p´^l f^*(t,s´) + k^l g^*(t,s´)) 
%\nonumber 
%\\
%&&( -i p^k f^*(t,s) + k^k g^*(t,s)) ( \sigma^k \sigma^i)_{s_f s_i} e^{i k (s-s´)} (\sigma^j \sigma^l)_{s_f s_i}
%\nonumber
%\\
%&&\hspace{3in}(-i p^l f(t,s´) + k^l g(t,s´)) \Bigg]
%\end{eqnarray}
%
where the functions $\Lambda$ denote,
\begin{eqnarray}
\Lambda^{l}(t,s) &=& ip^{l}f(t,s)+ k^{l}g(t,s)
\\
\Lambda'^{^{l}}(t,s') &=& ip'^{^{l}}f(t,s')+ k^{l}g(t,s')
\end{eqnarray}
with,
\begin{eqnarray}
f(t,s) &=& e^{-imt} + \cos m (t-2s)
\\
g(t,s) &=& e^{-im(t-s)} \sin m s
\end{eqnarray}
and $(\;)^{*}$ means complex conjugation.
Simplifications occur, when we ignore the spin degrees of freedom, that is we 
consider initial and final states of total ignorance for the spin. Hence we define 
the "spinless'' propagator 
\begin{eqnarray}
\lefteqn{K(x_f,y_f;t|x_i,y_i) = \int \frac{d^3{\bf p}}{(2 \pi)^3}
\frac{d^3{\bf p'}}{(2 \pi)^3} e^{i \big(p' (x_f - x_i) - p (y_f - y_i) 
-\frac{t}{2m} (p^2 - p'^2 )\big)}}\hspace{1.3in}
\nonumber
\\
&&\times\Bigg[ 1 + \frac{e^2}{m^2} \int \frac{d^3k}{(2 \pi)^3} e^{ik (x_i - y_i)} 
T^{ij}({\bf k}) \int_0^t ds \int_0^s ds'\hspace{0.2in}
\nonumber
\\
&&\hspace{0.5in}\times\Lambda^{k}(t,s) e^{-i k (s-s')} (\sigma^j \sigma^l)_{s_f s_i} 
(\Lambda'^{^{k}}(t,s'))^{*}\hspace{0.0in}
\nonumber 
\\
&&\hspace{0.53in}\times (\Lambda^{k}(t,s))^{*} e^{i k (s-s')} 
(\sigma^j \sigma^l)_{s_f s_i} \Lambda'^{^{l}}(t,s')\Bigg]\hspace{0.0in}
\end{eqnarray}
%
%Old Version
%
%\begin{equation}
%K(x_f,y_f;t|x_i,y_i) = \frac{1}{2} \delta^{r_f s_f}  \frac{1}{2} \delta^{r_i s_i} 
%K_{r_f r_i s_f s_i}(x_f,y_f;t|x_i,y_i)  
%\end{equation}
%which is found to be
%\begin{eqnarray} 
%K(x_f,y_f;t|x_i,y_i) = \int \frac{d^3{\bf p}}{(2 \pi)^3}
%\frac{d^3{\bf p'}}{(2 \pi)^3} e^{i p' (x_f - x_i) - i p (y_f - y_i) 
%-\frac{it}{2m} (p^2 - p'^2 )}
%\nonumber 
%\\
%\times \Bigg[ 1 + \frac{e^2}{m^2} \int \frac{d^3k}{(2 \pi)^3} e^{ik (x_i - y_i)}   
%\int_0^t ds \int_0^s ds´ 
%\nonumber 
%\\ 
%( i p^k f(t,s) + k^k g(t,s))  e^{-i k (s-s´)} (\sigma^j \sigma^l)_{s_f s_i}
%(-i p´^l f^*(t,s´) + k^l g^*(t,s´)) 
%\nonumber 
%\\
%( -i p^k f^*(t,s) + k^k g^*(t,s))  e^{i k (s-s´)} (\sigma^j \sigma^l)_{s_f s_i}
%(-i p^l f(t,s´) + k^l g(t,s´)) \Bigg]
%\end{eqnarray}
We now perform the ${\bf k}$ integration. Writing ${\bf r} = {\bf x}_i - {\bf y}_i$ we get
\begin{eqnarray}
\int \frac{d^3 k}{(2 \pi)^3} \frac{1}{2 k} e^{-i k r - ik s} = \frac{1}{4 \pi^2} \frac{1}{s^2 - r^2 } - \frac{i}{8 \pi r} \delta(s -r) = I(r,s)
\end{eqnarray}
In this integral the principal value is assumed for the real part, and a delta function on advanced time $s+r$ has been dropped out since we have positive $s$ in our integration range. 
\par
Similarly we can obtain
\begin{eqnarray}
\int \frac{d^3 k}{(2 \pi)^3} \frac{1}{2 k} e^{-i k r - ik s} k^i &=&  
i \frac{\partial}{\partial r_i} I(r,s) \\
\int \frac{d^3 k}{(2 \pi)^3} \frac{1}{2 k} e^{-i k r - ik s} k^2 &=& 
- \frac{\partial^2}{\partial r^2} I(r,s)
\end{eqnarray}
Using these expressions it is straightforward to perform the $p$ and $p´$ integration to obtain our final expression for the propagator
\begin{eqnarray}
\lefteqn{ K(x_f,y_f;t|x_i,y_i) = C(t) \exp \Bigg[-\frac{im}{2t} [(x_f - x_i)^2 
-(y_f -y_i)^2]}
\nonumber 
\\
&&+e^2 [ \alpha(t,r) (x_f - x_i)(y_f-y_i) 
\nonumber
\\
&&+{i \beta(t,r)\over r} (x_f - x_i) (x_i - y_i)
\nonumber
\\
&&-i\beta^*(t,r) (y_f - y_i)(x_i-y_i) - \gamma(t,r)] \Bigg]
\end{eqnarray}
The expressions for the coefficients $\alpha$, $\beta$, $\gamma$ are quite complicated. What we are more interested is their dependence on $r$.  Recalling that we are working in the 
non-relativistic regime and for the dimensions of our system $r$ can be taken much smaller than $t$. Hence to a first approximation one can expand the real part of $I(r,s)$ in a power series with respect to $r$ , which essentially corresponds to a power series with respect to $1/c$
(not so with the imaginary part due to the presence of the delta function). This gives for the coefficients  following $r$ dependence
\begin{eqnarray}
\alpha(t,r) &=& \alpha_1(t,r)/r + A(t) + O(r) \\
\beta(t,r) &=& \frac{\partial}{\partial r} b_1(t,r) + O(r)\\
\gamma(t,r) &=& \frac{\partial^2}{\partial r^2} \gamma_1(t,r) + O(r)
\end{eqnarray}
where $\alpha_1$,$\beta_1$, $\gamma_1$ denotes the integrals containing the delta function
which are calculable in terms of elementary trigonometric functions (see appendix).

\subsection{The Gaussian (Coulomb ) approximation}
 \par
Still, the $r$ dependence of the propagator is very complicated. Things simplify if one considers the "Coulomb" limit $c \rightarrow \infty$ or $r \rightarrow 0$. This essentially corresponds to ignoring the retarded propagation effects for the virtual photons and is expected to be a good approximation (at least for early times) when the spread of the wave packet is very small. Ignoring for the moment the domain of validity of this result,we can easily see that the functions $\beta$ and $\gamma$ vanish and $\alpha(t)$ contains only a time dependent contribution., hence the propagator becomes Gaussian in this limit
\begin{eqnarray}
K(x_f,y_f;t|x_i,y_i) = C(t) \exp \Bigg[ -\frac{im}{2t} [(x_f - x_i)^2 - (y_f -y_i)^2]\nonumber 
\\
+e^2  (\alpha(t) (x_f - x_i)(y_f-y_i) \Bigg] 
\end{eqnarray}
From this with the standard procedure one can  construct the master equation for
 the reduced density matrix of the "spinless'' particle:
\begin{eqnarray}
\frac{\partial}{\partial t} \rho(x,y) = - \frac{1}{2mi} \left( \frac{\partial^2}{\partial x^2} - \frac{\partial}{\partial y^2} \right) \rho(x,y) \nonumber \\
 \frac{2 e^2}{m^2} \frac{\dot{\alpha}t - \alpha}{t^2}
\frac{\partial^2}{\partial x \partial y} \rho(x,y)
\end{eqnarray}
or in operator form
\begin{eqnarray}
\frac{\partial}{\partial t} \hat{\rho} = \frac{1}{i} [\hat{H}  , \hat{\rho}] 
- \frac{2 e^2} 
{m^2} f(t) [\hat{p}, [ \hat{p}, \hat{\rho}]]
\end{eqnarray}
where
\begin{eqnarray}
f(t) = \frac{\dot{\beta}t - \beta}{t^2}
\end{eqnarray}
and its plot with time is given at the figure 1. We can see that it is oscillatory with a basic frequency of the order of $m^{-1}$.
\begin{figure}
\caption{The function $f(t)$, containing the effects of diffusion at times $t >> m^{-1}$.  } \label{u3}
 \centerline{%
   \psfig{file=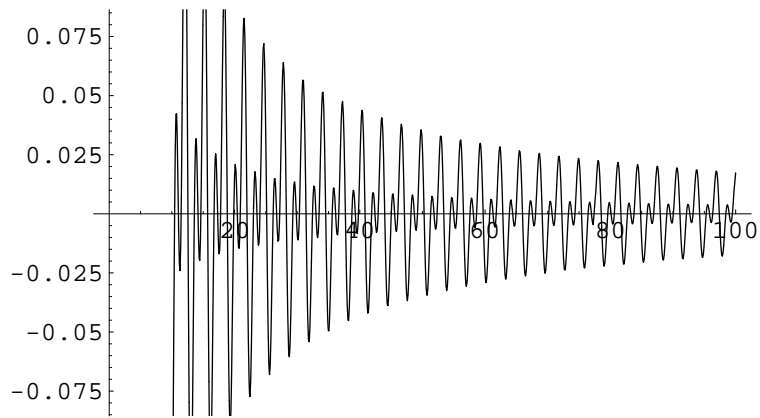,width=10cm,angle=0}%
   }
\end{figure}
   It is rather straightforward having this simple expression for the propagator, to examine whether in this regime there is any decoherence in position for the electron.                                            
    For this we consider the evolution of the initial state
\begin{equation}
\psi(x) = \left( \frac{a}{2 \pi} \right)^{1/4} \left( e^{-a x^2} + e^{-a (x - L)^2} \right)
\end{equation}
which is a superposition of two wavepackets, with their centers separated by  $L$.
Time evolution with the propagator (4.12) yields a mixed state with two diagonal terms propagating according to the classical equations of motion 
plus two off-diagonal terms. To examine the appearance of decoherence one is interested in computing the prefactor of the latter. This is found to be 
\begin{equation}
\exp \left( - \frac{a L^2 (\frac{m^2}{4 t^2} + e^4 \frac{\beta(t)^2}{t^4})}{a^2 + \frac{m^2}{4 t^2} + \frac{2 e^2}{t^2} \beta(t)} \right)
\end{equation}
It is easy to check that the off-diagonal terms are not suppressed. A necessary condition for that would be a growth of $\beta(t)$ of at least $t^2$ with time, while we can see from figure 
2, that $\beta(t)/t^2$ is falling asymptotically as $1/t$. So within this approximation , no decoherence in position seems to be possible for the electrons. This is in accordance with the facts of the classic two-slit experiment.
\begin{figure}
\caption{  The function $\beta(t)/t^2$ for $t >> m^{-1}$.  }
   \label{u1}
 \centerline{%
   \psfig{file=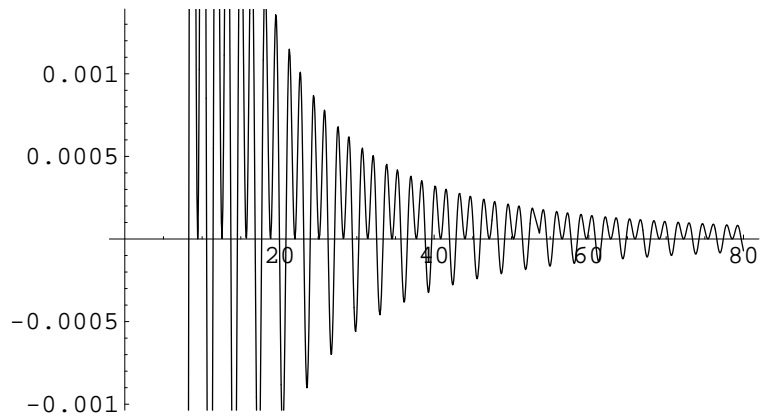,width=10cm,angle=0}%
   }
\end{figure}
\par
But how reliable is the approximation of switching off the $r$ dependence of the propagator, or rather in which regime should it be trusted. There are two remarks we should make. First, the term containing $\gamma(t,r)$ is increasing with time, hence at later times (depending primarily on the initial configuration of the system) this term becomes sufficiently large , to invalidate our perturbation scheme. On the other hand in the Gaussian approximation the perturbative terms in the propagator remain sufficiently small at all times. We therefore conclude, that as long as our Gaussian approximation is valid, perturbation theory is reliable.
Of course, both the Gaussian approximation and perturbation theory is due to break at the long time limit.
\par
What then can we say, about decoherence in position of the electron, outside this approximation. Unfortunately very little, because the term that we would expect to give rise to decoherence (namely $\gamma(t,r)$ ) is the one that eventually breaks perturbation theory and hence does not allow us to reach any general conclusions. The problem is aggravated, because the only natural length scale in our system is the Compton wavelength of the particle and qualitative arguments of length scale separation are not sufficient to give any insight.
\par
What we can conclude from our analysis is the following. There exists a regime dependent on the spread of the initial state of the electron, that one can safely use the Gaussian approximation. In such a regime no decoherence phenomena appear. As the spread of the initial state $L$ increases,  the influence functional phase gets a large real negative contribution which is characteristic of decoherence phenomena, and we are led to conjecture that decoherence is indeed possible.
\par
One can view these results in the light of the following qualitative 
observations. Decoherence in any system is connected with the propagation of coherent phase from the system to the environment. In our particular case, the carrier of the information is of course the photonic field. Hence decoherence is expected, when the configuration of the state of the particle is suitable to excite a number of photons , which propagate in the environment carrying the phase information. In the Gaussian regime as we have seen the flow of photons (essentially dissipation) is absent, while the $\beta$ term in the propagator- negligible in this case-
essentially contains the effects of dissipation.
\par
Now in the electron - photon vacuum system there are essentially two competing processes governing the emission of photons. On one hand the space where the electron wave function has support "contains" a charge distribution, which essentially acts as "screening" medium, that does not allow virtual photons to escape. The vacuum fluctuations here are of an essentially "Coulomb" type and can only give noise, whose effect only effect is  randomness in the evolution of the electron. This process is stronger when the size of the wavepacket is small, and it is precisely in this regime that our Gaussian approximation is valid.
\par
As the spread of the wavepacket increases, and in particular its shape deviates more from a spherical distribution another process starts gaining momentum. Higher moments of the charge distribution are coupled to the electromagnetic vacuum and as such they enhance photon emission. In particular, the case where the electron is in a superposition of two spatially localized states, the contribution from the dipole moment (rather than the quadrapole for more localized states) becomes significant and the electron essentially behaves as an oscillating dipole. When this effect becomes strong and sufficient to overcome screening, significant photon production occurs and decoherence is possible.
\par
At this point we should comment on the relation of our results with the existing bibliography.
 The density matrix propagator is distinctly different from the one obtained in \cite{BaCa}. 
These authors have been able to derive a Markovian limit of their master equation and obtain the 
Lorentz equation in the semiclassical limit. The crucial difference is that they have derived their 
result under the dipole approximation. For this they considered a restoring harmonic potential 
so that the particle is constrained to lie in a small spatial neighborhood. In our case having 
derived our master equation from full Quantum Electrodynamics, we 
have considered an essentially free electron. Since the only natural length scale appearing in our 
calculations is the Compton wavelength of the particle, and free evolution does not allow for 
localization of the particle, the dipole approximation should not be expected to be valid here.
\section{Conclusions}
The density matrix propagator (3.17) and (4.19) for
the single particle , we consider to be the most important results in
our paper. The former is to be seen as a starting point for many
interesting applications (to be outlined below), while the latter
provides (we believe) a conclusive demonstration that electromagnetic
field vacuum fluctuations are not sufficient to cause decoherence to
single 
free electrons.
\par  
It has been rather disappointing for us not to be yet able to say anything conclusive yet about 
the possibility of dynamically induced superselection rules. As we said earlier the complicated 
nature of our expressions preclude any possibility of being able to make any statement about 
generic states. Still we think we have provided the basic tools for a
more thorough investigation of these issues.
\par
The study of various regimes seems to be a natural step after this work. An important case can be
the case where the field lies either in the $N$ or the $M$ particle state. The propagator would 
then reads ${\bf P_{NM} J P_{NM}}$ where ${\bf P_{NM}}$ is defined by ${\bf P_{NM}} \rho = (P_N + P_M) \rho
(P_N + P_M)$, $P_N$ being the projector in the $N$ particle sector. If the hypothesis of the dynamical 
origin of charge superselection rule is valid then decoherence phenomena should be definitely 
clear in this regime.
\par
Another interesting issue towards an answer of which, our formalism can aim is the study of electrons in 
a background squeezed state. Squeezed states are characterized by negative energy densities, and
a connection between them and decoherence, dissipation and noise would be very rewarding.
\par
Finally, the expression (4.10) for the single electron propagator 
implies a spin - momentum self-coupling due to the vacuum
fluctuations. The study of this kind of couplings is a rather unexplored territory and potentially of much interest
in the light of quantum optics.

\begin{appendix}
\section{Coefficients in sigle-particle propagator}
We here give expressions for the functions $\alpha_1$, $\beta_1$, $\gamma_1$ and $A$ in equations (4.17 - 4.19).
\begin{eqnarray}
\alpha_1(t,r) &=& \frac{1}{4 \pi m} \sin^2 mt \sin^2mr \\
\beta_1(t,r) &=& \frac{1}{16 \pi r} \int_0^t ds \bigg( e^{-im(s-r)} 
\sin m(s-r) +\cos m(2t-s) e^{im(t-s+r)} 
\nonumber 
\\
&&- e^{ims} \sin ms  - \cos m(2t - s +r) e^{-im(t-s)} \bigg)
\\
\gamma_1(t,r) &=& \frac{\sin mr}{16 \pi m r} \left( t \cos mr 
+ \sin m(2t -r) + \sin mr \right)
\\
A(t) &=& \frac{1}{2 \pi^2} \int_0^t ds \int_0^s ds' \bigg( 1 + \cos 2m (s-s')
+ \cos 2m (t -s -s') 
\nonumber
\\  
&&+ \cos 2m (s-s') \cos 2m (s +s') 
\nonumber
\\
&&+ \cos 2m (s - s') \cos 2m (t - s -s') \bigg) 
\nonumber
\\
&& \times \frac{1}{(s - s')^2 + \epsilon^2}
\end{eqnarray}
Note, that the term $A$ has a cut-off dependence on $\epsilon$, which becomes smaller and at late times negligible. It can be absorbed into an early time redefinition according to our renormalisation scheme.
\par
Also, we can see that the term $\gamma_1$ is the one whose amplitude increases with time and is to be held responsible for the breakdown of our perturbative approximation and possibly as the one giving rise to decoherence.
\end{appendix}

\end{document}